\documentclass[aps,showpacs,preprint]{revtex4}

\begin{document}

\title{A Canonical Approach to the Einstein-Hilbert Action in Two Spacetime
Dimensions }
\author{N. Kiriushcheva, S.V. Kuzmin and D.G.C. McKeon}
\email{nkiriusc@uwo.ca,skuzmin2@uwo.ca,dgmckeo2@uwo.ca}

\affiliation{Department of Applied Mathematics, 
University of Western Ontario, London, N6A~5B7 Canada}

\date{\today}

\begin{abstract}
The canonical structure of the Einstein-Hilbert Lagrange density $L=\sqrt{-g}%
R$ is examined in two spacetime dimensions, using the metric density $h^{\mu
\nu }\equiv \sqrt{-g}g^{\mu \nu }$ and symmetric affine connection $\Gamma
_{\sigma \beta }^\lambda $ as dynamical variables. The Hamiltonian reduces
to a linear combination of three first class constraints with a local $%
SO(2,1)$ algebra. The first class constraints are used to find a generator
of gauge transformations that has a closed off-shell algebra and which
leaves the Lagrangian and $\det (h^{\mu \nu })$ invariant. These
transformations are distinct from diffeomorphism invariance, and are gauge 
transformations characterized by a symmetric matrix $\zeta_{\mu \nu}$.

\end{abstract}

\maketitle

The canonical structure of the $d$-dimensional Einstein-Hilbert (EH) action $%
S_d=\int d^dx\sqrt{-g}R$ has been examined for some time \cite{Dirac1958,ADM}%
. The two dimensional (2D) version of this action merits attention for the
insight it can provide, even though when expressed solely in terms of the
metric tensor $g_{\mu \nu }$ the Lagrangian reduces to a total derivative
and there are consequently no physical degrees of freedom. There has been
interest in analyzing the structure of this theory, despite its topological
nature \cite{Teitelboim,Labeotida,Montano}.

As dynamical variables, we select the metric density $h^{\mu \nu }\equiv 
\sqrt{-g}g^{\mu \nu }$ and
symmetric affine connection $\Gamma _{\sigma \beta }^\lambda $, as was done
originally by Einstein \cite{Ein1925} (though this is often called the
Palatini approach \cite{Pal}). We do not parameterize $h^{\mu \nu }$ in
a way that singles out the dynamics on a particular spatial surface, as
was done in \cite{Dirac1958,ADM}. Upon using these variables in 2D, we show
that it is particularly easy to apply the Dirac constraint formalism \cite
{Dirac1964} to analyze the canonical structure of the classical action $S_2$%
. Without having to even partially fix a gauge, we find that the
Hamiltonian reduces to a linear combination of three secondary first class
constraints. Unlike the constraints in the Dirac-ADM approach \cite
{Dirac1958,ADM} for $d > 2$ \cite{Teitelboim}, these constraints obey an 
algebra with field independent structure constants; it is a 
local $SO(2,1)$ algebra. A local algebraic structure has also been found 
in dilaton gravity but with field-dependent structure constants 
\cite{GrumKumVas2002}. A model of 2-dimensional gravity with an $SO(2,1)$ 
gauge symmetry appears in \cite{Cham}, though this model also involves 
an auxiliary scalar field. 

From the full set of first class constraints (both primary and secondary), a
generator of gauge transformations involving three local gauge parameters
can be constructed using the approach of Castellani \cite{Cast1982}. The
generator obeys a closed algebra, even off-shell, and results in a gauge
transformation that leaves $S_2$, $\det (h^{\mu \nu })$ and the equations of
motion invariant. The gauge transformation is distinct from the usual
diffeomorphism transformation.

The EH action in the first order formulation is

\begin{equation}
\label{1}S_d=\int d^dxh^{\alpha \beta }\left( \Gamma _{\alpha \beta ,\lambda
}^\lambda -\Gamma _{\alpha \lambda ,\beta }^\lambda +\Gamma _{\sigma \lambda
}^\lambda \Gamma _{\alpha \beta }^\sigma -\Gamma _{\sigma \beta }^\lambda
\Gamma _{\alpha \lambda }^\sigma \right) . 
\end{equation}

The Lagrange density of (\ref{1}) is polynomial of order three in the $\frac
12d\left( d+1\right) $ components of $h^{\mu \nu }$ and $\frac 12d^2\left(
d+1\right) $ components of $\Gamma _{\sigma \beta }^\lambda $ which are all
treated as being independent \cite{Ein1925}. This formulation is well suited
to a canonical analysis of $S_d$ because second order derivatives do not
appear at the outset in the action. There are special features associated 
with the first order formalism when $d = 2$ \cite{Gegenberg,Lindstrom,
Deser88,Deser95}. 
In this case, the metric obeys the constraint $det\left(h^{\mu \nu }\right)
 = -1$ and the equation of motion for $\Gamma _{\mu \nu }^\lambda $ does not 
uniquely fix it to be the Christoffel symbols $ \left\{_{\mu \nu }^\lambda 
\right\} = \frac{1}{2} g^{\lambda \sigma } \left( g_{\mu \sigma , \nu } + 
g_{\nu \sigma , \mu } - g_{\mu \nu , \sigma } \right)$.  

We introduce generalized momenta conjugate to all independent variables 
\begin{equation}
\label{4}\pi _{\alpha \beta }=\frac{\delta L}{\delta \dot h^{\alpha \beta }}%
,\Pi _\gamma ^{\alpha \beta }=\frac{\delta L}{\delta \dot \Gamma _{\alpha
\beta }^\gamma }. 
\end{equation}
Each of these equations constitutes a primary constraint

\begin{equation}
\label{10}\phi _{\alpha \beta }\equiv \pi _{\alpha \beta },\Phi _\gamma
^{\alpha \beta }\equiv \Pi _\gamma ^{\alpha \beta }-h^{\alpha \beta }\delta
_\gamma ^0+\frac 12\left( h^{\alpha 0}\delta _\gamma ^\beta +h^{\beta
0}\delta _\gamma ^\alpha \right) . 
\end{equation}

The total Hamiltonian $H_T$ is

\begin{equation}
\label{11}H_T=H_c+\lambda ^{\alpha \beta }\phi _{\alpha \beta }+\Lambda
_{\alpha \beta }^\gamma \Phi _\gamma ^{\alpha \beta }, 
\end{equation}
where $\lambda ^{\alpha \beta }$ and $\Lambda _{\alpha \beta }^\gamma $ are
undetermined multipliers and $H_c$ is the usual canonical Hamiltonian given
by 
\begin{equation}
\label{17}H_c=-h^{\alpha \beta }\Gamma _{\alpha \beta ,k}^k+h^{\alpha
k}\Gamma _{\alpha \lambda ,k}^\lambda -h^{\alpha \beta }\Gamma _{\sigma
\lambda }^\lambda \Gamma _{\alpha \beta }^\sigma +h^{\alpha \beta }\Gamma
_{\sigma \beta }^\lambda \Gamma _{\alpha \lambda }^\sigma . 
\end{equation}
(Latin indices are spatial.) The fundamental Poisson brackets (PB) for
canonical variables are

\begin{equation}
\label{18}\left\{ h^{\mu \nu },\pi _{\alpha \beta }\right\} =\Delta _{\alpha
\beta }^{\mu \nu },\left\{ \Gamma _{\alpha \beta }^\gamma ,\Pi _\lambda
^{\sigma \rho }\right\} =\delta _\lambda ^\gamma \Delta _{\alpha \beta
}^{\sigma \rho }, 
\end{equation}
where $\Delta _{\alpha \beta }^{\sigma \rho }=\frac 12\left( \delta _\alpha
^\sigma \delta _\beta ^\rho +\delta _\beta ^\sigma \delta _\alpha ^\rho
\right) $.

Time independence of the primary constraints of (\ref{10}) can either lead
to new (secondary) constraints or fixing of some the multipliers in (\ref{11}%
). The rank of the matrix constructed from the PB of primary constraint

\begin{equation}
\label{13}M=\left( \left\{ \phi _n,\phi _m\right\} \right) 
\end{equation}
(where $\phi _n=\left( \phi _{\alpha \beta },\Phi _c^{ab}\right) $)
corresponds to the number of multipliers that are determined and the
independent eigenvectors with zero eigenvalue produce the secondary
constraints. The only non-zero PB among the primary constraints is

\begin{equation}
\label{12}\left\{ \phi _{\alpha \beta },\Phi _\gamma ^{\sigma \rho }\right\}
=\Delta _{\alpha \beta }^{\sigma \rho }\delta _\gamma ^0-\frac 12\left(
\Delta _{\alpha \beta }^{\sigma 0}\delta _\gamma ^\rho +\Delta _{\alpha
\beta }^{\rho 0}\delta _\gamma ^\sigma \right) . 
\end{equation}

From the matrix of (\ref{13}) one can now find the secondary constraints.
These secondary constraints must also be time independent; this may imply
further tertiary constraints.

In the case $d=2$ the canonical analysis simplifies considerably as $H_c$ in
this instance is linear in variables whose associated momenta are first
class primary constraints. (For $d>2,$ the dependance on such variables is
non-linear, complicating the analysis considerably by leading to tertiary
constraints.) We have found that for 
$d=2$, the Dirac analysis shows that there are only primary and secondary
constraints, and of these, six are first class and six second class. (When
combined with six gauge conditions, this serves to eliminate all eighteen
canonical degrees of freedom associated with $h^{\mu \nu }$, $\Gamma
_{\sigma \beta }^\lambda $ and their conjugate momenta).

After having made this analysis, second class constraints of the form $%
p_i\simeq 0,$ $q_i\simeq f_i\left( p_{j\neq i},q_{j\neq i}\right) $ can be
used to explicitly eliminate the degrees of freedom associated with $p_i$
and $q_i$ by setting $p_i=0$ and $q_i$ equal to $f_i\left( p_{j\neq
i},q_{j\neq i}\right) $ in the Hamiltonian and remaining constraints \cite
{Dirac1964}; this simplifies our analysis of gravity in two or more
dimensions.

Equation (\ref{10}) gives nine primary constraints

\begin{equation}
\label{14a}\phi _{00}=\pi _{00},\phi _{01}=\pi _{01},\phi _{11}=\pi _{11}, 
\end{equation}

\begin{equation}
\label{14b}\Phi _0^{11}=\Pi _0^{11}-h^{11},\Phi _0^{01}=\Pi _0^{01}-\frac
12h^{01},\Phi _1^{01}=\Pi _1^{01}+\frac 12h^{00}, 
\end{equation}

\begin{equation}
\label{14}\Phi _0^{00}=\Pi _0^{00},\Phi _1^{00}=\Pi _1^{00},\Phi _1^{11}=\Pi
_1^{11}+h^{01}. 
\end{equation}

The matrix (\ref{13}) for the primary constraints (\ref{14a}-\ref{14}) has
rank six; hence there are six second class constraints (those of (\ref{14a},%
\ref{14b}) which we group into three pairs of the special form $\left( \phi
_{00},\Phi _1^{01}\right) ,\left( \phi _{11},\Phi _0^{11}\right) $ and $%
\left( \phi _{01},\Phi _0^{01}\right) $ (the last pair could also be taken
to be $\left( \phi _{01},\Phi _1^{11}\right) $)). Elimination of these
constraints by setting $\pi _{00}=\pi _{01}=\pi _{11}=0$ and $h^{11}=\Pi
_0^{11},h^{01}=2\Pi _0^{01},h^{00}=-2\Pi _1^{01}$ converts the Hamiltonian (%
\ref{17}) and the remaining constraints (\ref{14}) into

\begin{equation}
\label{32}H_T^{\left( 1\right) }=H_c^{\left( 1\right) }+\Lambda \Phi
+\Lambda ^1\Phi _1+\Lambda _1\Phi ^1, 
\end{equation}
where

\begin{equation}
\label{33a}\Phi \equiv \Phi _0^{00};\Phi _1\equiv \Phi _1^{00};\Phi ^1\equiv
\Phi _1^{11}=2\Pi _0^{01}+\Pi _1^{11} 
\end{equation}
and, after some rearrangement of terms,

$$
H_c^{\left( 1\right) }=\left( \Gamma _{01}^1-\Gamma _{00}^0\right) \left(
2\Pi _{0,1}^{01}+\Pi _0^{11}\Gamma _{11}^0-2\Pi _1^{01}\Gamma _{01}^1\right) 
$$

\begin{equation}
\label{22}-\Gamma _{00}^1\left( 2\Pi _{1,1}^{01}+2\Pi _1^{01}\left( \Gamma
_{01}^0-\Gamma _{11}^1\right) -4\Pi _0^{01}\Gamma _{11}^0\right) -\Gamma
_{01}^0\left( \Pi _{0,1}^{11}-\Pi _0^{11}\left( \Gamma _{01}^0-\Gamma
_{11}^1\right) +4\Pi _0^{01}\Gamma _{01}^1\right) .
\end{equation}

The phase space now consists of only $\Gamma $'s and their corresponding
momenta $\Pi $'s with the standard fundamental PB (\ref{18}). As the time
derivative of the primary constraints (\ref{33a}) must vanish we obtain
three secondary constraints

\begin{equation}
\label{40}\dot \Phi =\left\{ \Phi ,H_c^{\left( 1\right) }\right\} =2\Pi
_{0,1}^{01}+\Pi _0^{11}\Gamma _{11}^0-2\Pi _1^{01}\Gamma _{01}^1\equiv \chi
, 
\end{equation}

\begin{equation}
\label{41}\dot \Phi _1=2\Pi _{1,1}^{01}+2\Pi _1^{01}\left( \Gamma
_{01}^0-\Gamma _{11}^1\right) -4\Pi _0^{01}\Gamma _{11}^0\equiv \chi _1, 
\end{equation}

\begin{equation}
\label{42}\dot \Phi ^1=\Pi _{0,1}^{11}-\Pi _0^{11}\left( \Gamma
_{01}^0-\Gamma _{11}^1\right) +4\Pi _0^{01}\Gamma _{01}^1\equiv \chi ^1, 
\end{equation}

This converts the Hamiltonian (\ref{22}) into linear combination of
secondary constraints.

All primary constraints $C_P^a=\left( \Phi ,\Phi _1,\Phi ^1\right) $ have a
zero PB among themselves and with the secondary constraints $C_S^a=\left(
\chi ,\chi _1,\chi ^1\right) $. The only non-zero PB among the constraints
are local:

\begin{equation}
\label{33}\left\{ \chi (x) ,\chi _1 (y)\right\} =-\chi _1 \delta \left( x-y
\right),\left\{ \chi (x),\chi^1 (y) \right\} =\chi ^1 \delta \left( x-y
\right),\left\{ \chi ^1 (x),\chi _1 (y) \right\} =2\chi \delta \left( x-y
\right). 
\end{equation}

There are no tertiary constraints. After the redefinition 
$$
\sigma _a=\frac 12\left( \chi ^1+\chi _1\right) ,\sigma _b=\frac 12\left(
\chi ^1-\chi _1\right) ,\sigma _c=\chi 
$$
the algebra of (\ref{33}) becomes

$$
\left\{ \sigma _a,\sigma _b\right\} =\sigma _c,\left\{ \sigma _c,\sigma
_a\right\} =\sigma _b,\left\{ \sigma _b,\sigma _c\right\} =-\sigma _a 
$$

Upon replacing the classical PB by $\left( -i\right) \times $ (quantum 
commutator), this becomes the Lie algebra of $SO\left(2,1\right)$.

The approach of Castellani \cite{Cast1982} can be used to find the form of
the gauge transformations implied by the six first class constraints. The
generator $G$ is found by first setting $G_{\left( 1\right) }^a=C_P^a$ and
then examining $G_{\left( 0\right) }^a\left( x\right) =-C_S^a\left( x\right)
+\int dy\ \alpha _b^a\left( x,y\right) C_P^b\left( y\right) .$ The functions 
$\alpha _b^a\left( x,y\right) $ are found by requiring that $\left\{
G_{\left( 0\right) }^a,H_c\right\} $ vanish when the primary constraints
vanish. The generator of the gauge transformation is given by $G\left(
\varepsilon ^a,\dot \varepsilon ^a\right) =\int dx\left( \varepsilon
^a\left( x\right) G_{\left( 0\right) }^a\left( x\right) +\dot \varepsilon
^a\left( x\right) G_{\left( 1\right) }^a\left( x\right) \right) $, which in
our case leads to the following expression

$$
G\left( \varepsilon \right) \equiv G\left( \varepsilon ,\dot \varepsilon
;\varepsilon _1,\dot \varepsilon _1;\varepsilon ^1,\dot \varepsilon
^1\right) =\int dx\left[ \varepsilon \left( -\chi +\Gamma _{01}^1\Phi
+\Gamma _{00}^1\Phi _1-\Gamma _{01}^0\Phi ^1+\Phi _{,1}^1\right) +\dot
\varepsilon \Phi \right.  
$$

$$
+\varepsilon _1\left( -\chi ^1-2\Gamma _{00}^1\Phi +\left( \Gamma
_{00}^0+\Gamma _{01}^1\right) \Phi ^1\right) +\dot \varepsilon _1\Phi ^1 
$$

\begin{equation}
\label{50}\left. +\varepsilon ^1\left( -\chi _1+\left( \Gamma _{01}^0+\Gamma
_{11}^1\right) \Phi -\Phi _{,1}-\left( \Gamma _{00}^0-\Gamma _{01}^1\right)
\Phi _1\right) -\varepsilon ^12\Gamma _{11}^0\Phi ^1+\dot \varepsilon ^1\Phi
_1\right] .
\end{equation}

The PB of these generators have an algebra that closes off shell 
\begin{equation}
\label{51}\left\{ G\left( \varepsilon \right) ,G\left( \eta \right) \right\}
=G\left( \zeta ^a=C^{abc}\varepsilon ^b\eta ^c\right) 
\end{equation}

where $\varepsilon ^a=\left( \varepsilon ,\varepsilon _1,\varepsilon
^1\right) $, $\eta ^a=\left( \eta ,\eta _1,\eta ^1\right) $, $\zeta
^a=\left( \zeta ,\zeta _1,\zeta ^1\right) $ and the only non-zero structure
functions are $C^{123}=2=-C^{132}$, $C^{213}=1=-C^{231}$, $%
C^{321}=1=-C^{312} $.

If we compute the gauge transformation of fields by taking their PB with the
generator $G$, then the Lagrangian is invariant under these transformations
only when the constraints themselves vanish and the fields are on shell as
in the Dirac-ADM formulation of gravity \cite{Cast1982}. However, by a
contact transformation that correponds to a slight change in the choice of
dynamical variables in the 2D EH action, it is possible to find gauge
transformations that leave the Lagrangian invariant even off the constraint
surface and when fields are off shell. We make a linear change of variables,
suggested by (\ref{22}),

\begin{equation}
\label{53}\xi =\Gamma _{01}^1-\Gamma _{00}^0,\xi ^1=\Gamma _{00}^1,\xi
_1=\Gamma _{01}^0,\Sigma _{11}=\Gamma _{11}^0,\Sigma _1=\Gamma
_{01}^0-\Gamma _{11}^1,\Sigma =\Gamma _{01}^1,
\end{equation}
so that the Lagrangian becomes

\begin{equation}
\label{54}L=h^{11}\dot \Sigma _{11}+h^{01}\dot \Sigma _1-h^{00}\dot \Sigma
-H_c
\end{equation}
with 
$$
H_c=\xi \left( h_{,1}^{01}+h^{11}\Sigma _{11}+h^{00}\Sigma \right) +\xi
^1\left( h_{,1}^{00}+h^{00}\Sigma _1+2h^{01}\Sigma _{11}\right) -\xi
_1\left( h_{,1}^{11}-h^{11}\Sigma _1+2h^{01}\Sigma \right) . 
$$

We can now repeat the Dirac procedure starting from (\ref{54}). Introducing
momenta for the variables (\ref{53}) with the non-vanishing PB $\left\{ \xi
,\pi \right\} =\left\{ \xi ^1,\pi _1\right\} =\left\{ \xi _1,\pi ^1\right\}
=\left\{ \Sigma ,\Pi \right\} $ $=\left\{ \Sigma _1,\Pi ^1\right\} =\left\{
\Sigma _{11},\Pi ^{11}\right\} =1$ we obtain nine primary constraints, those
of (\ref{14a}), and in addition,

\begin{equation}
\label{57}\pi =0,\pi _1=0,\pi ^1=0, 
\end{equation}

\begin{equation}
\label{58}\Pi ^{11}-h^{11}=0,\Pi ^1-h^{01}=0,\Pi +h^{00}=0. 
\end{equation}

As before, after the elimination of the primary second class constraints (%
\ref{14a}) and (\ref{58}) we obtain the total Hamiltonian in reduced phase
space with the three primary constraints (\ref{57}),

\begin{equation}
\label{60}H_T=H_c+\lambda \pi +\lambda ^1\pi _1+\lambda _1\pi ^1
\end{equation}
where 

\begin{equation}
\label{61}H_c=\xi \left( \Pi _{,1}^1+\Pi ^{11}\Sigma _{11}-\Pi \Sigma
\right) +\xi ^1\left( -\Pi _{,1}-\Pi \Sigma _1+2\Pi ^1\Sigma _{11}\right)
-\xi _1\left( \Pi _{,1}^{11}-\Pi ^{11}\Sigma _1+2\Pi ^1\Sigma \right) .
\end{equation}

The time derivatives of the remaining primary constraints vanish if we have
the secondary constraints

\begin{equation}
\label{62}-\left( \Pi _{,1}^1+\Pi ^{11}\Sigma _{11}-\Pi \Sigma \right)
\equiv \tilde \chi , 
\end{equation}

\begin{equation}
\label{63}-\left( -\Pi _{,1}-\Pi \Sigma _1+2\Pi ^1\Sigma _{11}\right) \equiv
\tilde \chi _1, 
\end{equation}

\begin{equation}
\label{64}+\left( \Pi _{,1}^{11}-\Pi ^{11}\Sigma _1+2\Pi ^1\Sigma \right)
\equiv \tilde \chi ^1 
\end{equation}
whose algebra is

\begin{equation}
\label{65}\left\{ \tilde \chi ,\tilde \chi _1\right\} =\tilde \chi
_1,\left\{ \tilde \chi ,\tilde \chi ^1\right\} =-\tilde \chi ^1,\left\{
\tilde \chi ^1,\tilde \chi _1\right\} =-2\tilde \chi . 
\end{equation}

This algebra is identical to (\ref{33}) and ensures that the time
derivatives of all secondary constraints weakly vanish. The gauge generator
becomes 
$$
\tilde G\left( \varepsilon \right) =\int dx\left[ +\varepsilon \left(
-\tilde \chi -\xi ^1\pi _1+\xi _1\pi ^1\right) +\dot \varepsilon \pi \right.
$$

\begin{equation}
\label{67}\left. +\varepsilon _1\left( -\tilde \chi ^1+2\xi ^1\pi -\xi \pi
^1\right) +\dot \varepsilon _1\pi ^1+\varepsilon ^1\left( -\tilde \chi
_1-2\xi _1\pi +\xi \pi _1\right) +\dot \varepsilon ^1\pi _1\right] .
\end{equation}

This too satisfies the algebra (\ref{51}). It can be shown that now $\left\{
\tilde G_{\left( 0\right) }^a,H_c\right\} =0$ as for Yang-Mills theory \cite
{Cast1982}. Consequently, the variables (\ref{53}) lead to a local algebra
of constraints with field independent structure constants and a closed
algebra of generators off the constraint surface and off shell that
generates transformations which leave $S_2$ invariant 
off shell. We thus have a truly canonical formulation of the 2D EH action.

$\tilde G$ gives rise to the transformations 
$$
\delta \pi =\varepsilon _1\pi ^1-\varepsilon ^1\pi _1,\delta \pi
_1=\varepsilon \pi _1-2\varepsilon _1\pi ,\delta \pi ^1=-\varepsilon \pi
^1+2\varepsilon ^1\pi , 
$$

$$
\delta \Pi =\varepsilon \Pi +2\varepsilon _1\Pi ^1,\delta \Pi
^1=-\varepsilon _1\Pi ^{11}+\varepsilon ^1\Pi ,\delta \Pi ^{11}=-\varepsilon
\Pi ^{11}-2\varepsilon ^1\Pi ^1, 
$$

$$
\delta \xi =\dot \varepsilon +2\varepsilon _1\xi ^1-2\varepsilon ^1\xi
_1,\delta \xi _1=\dot \varepsilon _1+\varepsilon \xi _1-\varepsilon _1\xi
,\delta \xi ^1=\dot \varepsilon ^1-\varepsilon \xi ^1+\varepsilon ^1\xi , 
$$

\begin{equation}
\label{68}\delta \Sigma =\varepsilon _{,1}^1-\varepsilon \Sigma
-\varepsilon ^1\Sigma _1,\delta \Sigma _1=-\varepsilon _{,1}-2\varepsilon
_1\Sigma +2\varepsilon ^1\Sigma _{11},\delta \Sigma _{11}=\varepsilon
_{1,1}+\varepsilon \Sigma _{11}+\varepsilon _1\Sigma _1
\end{equation}

which by (\ref{58}) and (\ref{53}) gives

$$
\delta h^{00}=\varepsilon h^{00}-2\varepsilon _1h^{01},\delta
h^{01}=-\varepsilon _1h^{11}-\varepsilon ^1h^{00},\delta h^{11}=-\varepsilon
h^{11}-2\varepsilon ^1h^{01}, 
$$

$$
\delta \Gamma _{11}^0=\varepsilon _{1,1}+\varepsilon \Gamma
_{11}^0+\varepsilon _1\left( \Gamma _{01}^0-\Gamma _{11}^1\right) ,\delta
\Gamma _{01}^1=\varepsilon _{,1}^1-\varepsilon \Gamma _{01}^1-\varepsilon
^1\left( \Gamma _{01}^0-\Gamma _{11}^1\right) , 
$$

$$
\delta \Gamma _{11}^1=\dot \varepsilon _1+\varepsilon _{,1}+\varepsilon
\Gamma _{01}^0+\varepsilon _1\left( \Gamma _{01}^1+\Gamma _{00}^0\right)
-2\varepsilon ^1\Gamma _{11}^0, 
$$

$$
\delta \Gamma _{01}^0=\dot \varepsilon _1+\varepsilon \Gamma
_{01}^0-\varepsilon _1\left( \Gamma _{01}^1-\Gamma _{00}^0\right) ,\delta
\Gamma _{00}^1=\dot \varepsilon ^1-\varepsilon \Gamma _{00}^1+\varepsilon
^1\left( \Gamma _{01}^1-\Gamma _{00}^0\right) , 
$$

\begin{equation}
\label{69}\delta \Gamma _{00}^0=-\dot \varepsilon +\varepsilon
_{,1}^1-\varepsilon \Gamma _{01}^1-\varepsilon ^1\left( \Gamma
_{01}^0+\Gamma _{11}^1\right) -2\varepsilon _1 \Gamma _{00}^1. 
\end{equation}

Using these transformations it is straightforward calculation to demonstrate
that under (\ref{69}) the 2D action (\ref{1}) is invariant, $\delta S_2=0$.
This is an exact result as in Yang-Mills theory (not just on the constraint
surface or on shell as in the case of generators of general coordinate 
transformations in gravity, derived from the Dirac-ADM  constraint analysis 
of the EH action \cite{Cast1982}). The transformation of (\ref{69}) 
is distinct from a diffeomorphism, which is immediately apparent as it is 
characterized by three rather than two parameters. Indeed, (\ref{69}) can be 
rewritten in a way that resembles the transformations appearing in 
\cite{Deser88} if we use the antisymmetric tensor 
$\epsilon^{\alpha \beta }$ and affine covariant derivatives $D_\rho 
\zeta_{\mu \nu} = \partial_\rho \zeta_{\mu \nu} - \Gamma _{\rho \mu}^\sigma 
\zeta_{\sigma \nu} - \Gamma _{\rho \nu}^\sigma \zeta_{\sigma \mu}$:

\begin{equation}
\label{70} \delta h^{\alpha \beta } = -\left(\epsilon^{\alpha \lambda } 
 h^{\sigma \beta } +\epsilon^{\beta \lambda } h^{\sigma \alpha } \right) 
\zeta_{\lambda \sigma }
\end{equation}

\begin{equation}
\label{71} \delta \left[ \Gamma_{\mu \nu }^\lambda - \frac{1}{2} \left(
\delta^\lambda_\mu \Gamma_{\nu \sigma}^\sigma + \delta^\lambda_\nu 
\Gamma_{\mu \sigma}^\sigma \right) \right] = \epsilon^{\lambda \rho} 
D_\rho \zeta_{\mu \nu } + \epsilon ^{\lambda \rho } \Gamma_{\rho \sigma}^
\sigma \zeta_{\mu \nu },
\end{equation}

where $\zeta_{00} = -\varepsilon^1,\zeta_{11} = \varepsilon_1$ and 
$\zeta_{01} = \zeta_{10} = -\frac{1}{2}\varepsilon$.   

We note that as in $d$ dimensions, $\det (h^{\mu \nu })=-(-g)^{d/2-1}$, and
hence in 2D, there is the extra condition 
\begin{equation}
\label{X1}\Xi ^{11}\equiv h^{00}h^{11}-(h^{01})^2=-1. 
\end{equation}
However, with the generator $\tilde G$ (\ref{67}), it is easy to show that $%
\left\{ \tilde G,\Xi ^{11}\right\} =0$ and hence our formalism is consistent
with (\ref{X1}). Supplementing the action $S_2$ with a term $\lambda \left(
h^{00}h^{11}-(h^{01})^2+\rho\right) $ leads to a pair of primary constraints 
$p_\lambda = p_\rho = 0$ where $p_\lambda$ and $p_\rho$ are momenta 
conjugate to $\lambda$ and $\rho$ respectively. There then follows the 
secondary constraint $\Xi ^{11}_\rho \equiv \Xi ^{11} + \rho = 0$. 
The constraint 
$p_\lambda = 0$ can be associated with the gauge condition $\lambda = 0$; 
 $p_\rho = 0$ and $\Xi ^{11}_\rho = 0$ are second class constraints. Thus, 
there is no net change in the number of degrees of freedom in the model as 
a result of imposing the condition $\Xi ^{11}_\rho = 0$ - there are still zero 
degrees of freedom left after all constraints are applied. (In some models 
of two dimensional gravity, there are a negative number of degrees of 
freedom \cite{Martinec,Polchinski}.)

If instead of using $h^{\mu \nu }$ as a dynamical field, we were to use the
metric $g^{\mu \nu }$, then it would not be necessary to impose (\ref{X1}).
As will be reported elsewhere, the use of $g^{\mu \nu }$ in place of $h^{\mu
\nu }$ results in there being seven first class constraints and gauge
generator involving five independent functions, though the all canonical
properties (such as having a local algebra and possessing off shell
invariance, etc.) are the same as in the formalism developed above.

Application of the procedure employed in this letter to the EH action in $%
d>2 $ dimensions is being considered; it is expected that the simple
canonical properties present in ordinary gauge theories could be preserved
in these dimensions as well. This would provide an interesting alternative
approach to quantum gravity.

This work is supported in part by funds provided by NSERC.

D.G.C.McKeon would like to thank Perimeter Institute for hospitality while
part of this work was completed and R. and D. MacKenzie for helpful advice.

\end{document}